\documentclass[a4paper,12pt]{article}
\usepackage[dvips]{graphicx}
\begin{document}
\begin{center}
\Large\bf
Time evolution of relativistic d + Au and Au + Au collisions\\[2.1cm]
\large\rm
Georg Wolschin$^{1}$, Minoru
Biyajima$^{2}$,\\ Takuya Mizoguchi$^{3}$, and Naomichi Suzuki$^{4}$\\[.8cm]
\normalsize\sc\rm
$^{1}$ Institut f\"ur Theoretische Physik der Universit\"at, 
D-69120 Heidelberg, Germany\\
$^{2}$ Department of Physics, Shinshu University, Matsumoto 390-8621,
Japan\\
$^{3}$ Toba National College of Maritime Technology, Toba 517-8501, Japan\\
$^{4}$ Department of Comprehensive Management,\\ Matsumoto University,
Matsumoto 390-1295,
Japan\\[2.6cm]
\end{center}
\rm
The evolution of charged-particle production in collisions of heavy 
ions at relativistic energies is investigated as function of 
centrality in a nonequilibrium-statistical framework. Precise 
agreement with recent d + Au and Au + Au data at $\sqrt{s_{NN}}$ = 200 GeV
is found in a 
Relativistic Diffusion Model with three sources for particle
production. Only the midrapidity source 
comes very close to local equilibrium, whereas the analyses 
of the overall pseudorapidity
distributions show that the systems remain far from statistical
equilibrium.
\\[.4cm]
\newpage
\section{Introduction}
The investigation of particle production in relativistic heavy-ion collisions
at the highest available energies offers an ideal opportunity to 
study the gradual approach to
statistical equilibrium in a strongly interacting many-particle
system. It has recently been shown by several groups that
analytically soluble non-equilibrium statistical models are suitable
to accurately describe a fairly large amount of phenomena that are
observed experimentally. This allows us to determine
how closely the system approaches thermal equilibrium in the course
of particle production during the collision.

In particular, pseudorapidity
distributions of primary charged particles have become available
\cite{bbb05} as functions of centrality in d + Au collisions at
a nucleon-nucleon center-of-mass energy of 200 GeV. They complement
corresponding data on particle production for the heavy 
Au + Au system at various incident energies \cite{bb105,nou02}. Both are investigated
within a nonequilibrium-statistical framework that is based on
analytical solutions of a Relativistic Diffusion Model (RDM).

We investigate these solutions as functions of time for both 
asymmetric, and symmetric systems in comparison with the data. 
Whereas the midrapidity source 
for particle production comes very close to equilibrium in rapidity space, 
this is not the case for
the target- and projectile-like sources. They remain far from 
equilibrium, and produce the characteristic nonequilibrium shape of
the overall rapidity distribution function. This
is particularly evident in case of the asymmetric d + Au system,
where the steeper slope in the deuteron direction is shown to be an immediate 
consequence of the nonequilibrium properties. A short account of this
work has been given in \cite{wbs05}.

The analytical model is outlined in section 2. The time dependence of 
the solutions in pseudorapidity space, and the pseudorapidity 
distributions for produced charged hadrons as functions of collision 
centrality are obtained in section 3 for d + Au, and in section 4 for
Au + Au. The conclusions are drawn in section 5.
\newpage
\section{Relativistic Diffusion Model}

Nonequilibrium processes such as those observed in the
course of particle production during relativistic heavy-ion collisions
have successfully been described in many areas of physics by
Fokker-Planck equations (FPEs) \cite{fok14,kam97}.
These were also used in Brownian motion of macroscopic particles
in a heat bath \cite{ray91,ein05,uhl30}, where they allow
to model both the velocity- and the spatial distribution
of test particles. Due to the large number of produced particles in
relativistic heavy-ion collisions and the random nature
of their mutual strong interactions \cite{rhic05}, such equations are
useful in the detailed modelling of the distribution
functions even though there is no heat bath present. However, 
Lorentz-invariant kinematical variables have to be introduced. 
In particular, the rapidity replaces the velocity to describe the motion 
parallel to the beam direction, and for the transverse motion
a corresponding transverse rapidity may be introduced.

In this work we concentrate on the ordinary
(longitudinal) rapidity. The analysis will show that thermal
equilibrium is not reached in particle production,
but one comes sufficiently close to it to justify the use of the FPE.
The present investigation is based on a linear  
Fokker-Planck equation 
for three components $R_{k}(y,t)$ of the distribution function
for produced charged hadrons in rapidity space
\cite{wol99,biy02,wol03} 

\begin{equation}
\frac{\partial}{\partial t}R_{k}(y,t)=
\frac{1}{\tau_{y}}\frac{\partial}
{\partial y}\Bigl[(y-y_{eq})\cdot R_{k}(y,t)\Bigr]
+\frac{\partial^2}{\partial^{2} y}\Bigl[D_{y}^{k}
\cdot R_{k}(y,t)\Bigr]
\label{fpe}
\end{equation}\\
with the rapidity $y=0.5\cdot ln((E+p)/(E-p))$.
The diagonal components $D_{y}^{k}$ of the diffusion tensor  
contain the microscopic
physics in the respective target-like (k=1), projectile-like (k=2)
and central (k=3) regions. They 
account for the broadening of the distribution 
functions through interactions and particle creations. 
In the present investigation the off-diagonal terms of the
diffusion tensor are assumed to be zero.
The rapidity relaxation time $\tau_{y}$ determines
the speed of the statistical equilibration in y-space.

As time goes to infinity, the mean values of the
solutions of Eqs. (\ref{fpe}) approach the equilibrium value $y_{eq}$. 
We determine it 
from energy- and momentum conservation \cite{bha53,nag84}
in the system of target- and projectile-participants and hence, it 
depends on impact parameter. This dependence is decisive 
for a detailed description of the measured charged-particle
distributions in asymmetric systems:

\begin{equation}
y_{eq}(b)=1/2\cdot ln\frac{<m_{1}^{T}(b)>exp(y_{max})+<m_{2}^{T}(b)>
exp(-y_{max})}
{<m_{2}^{T}(b)>exp(y_{max})+<m_{1}^{T}(b)>exp(-y_{max})}
\label{yeq}
\end{equation}\\
with the beam rapidities y$_{b} = \pm y_{max}$, the transverse
masses $<m_{1,2}^{T}(b)>=\\
\sqrt(m_{1,2}^2(b)+<p_{T}>^2)$, and masses
m$_{1,2}(b)$ of the target- and projectile-like participants 
that depend on the impact parameter $b$. The average 
numbers of participants $<N_{1,2}(b)>$
in the incident nuclei are calculated from the
geometrical overlap. The results are consistent with the Glauber
calculations reported in \cite{bbb05} for d + Au and in \cite{bb105}
for Au + Au which we use in the further
analysis. 

The corresponding equilibrium values of the rapidity are zero in
symmetric systems. In the asymmetric d + Au case, they
vary from y$_{eq}=$ - 0.169 for peripheral (80-100$\%$) to 
y$_{eq}=$ - 0.944 for central (0-20$\%$) collisions.
They are negative due to the net longitudinal momentum of the
participants in the laboratory frame, and their absolute
magnitudes decrease with increasing impact parameter since the number of
participants decreases for more peripheral collisions.

The RDM describes the drift of the mean values of the
partial distributions towards $y_{eq}$. The existence
of this drift has clearly been established from the comparison of 
RDM-results with net-proton rapidity distributions at
various incident energies \cite{wol03}, where it is directly
visible in the available data from the NA 49 and BRAHMS collaborations.
For produced hadrons, the drift of the partial distribution functions
is not directly visible in the data, although its presence
is essential for a precise modeling of the results.

Whether the mean values of the distribution functions $R_{1}$ and 
$R_{2}$ actually attain $y_{eq}$ depends on the interaction time 
$\tau_{int}$ (the time the system interacts strongly, or the integration time
of (\ref{fpe})). It can be determined from dynamical models or
from parametrizations of two-particle correlation measurements. For
central Au + Au at 200 A GeV, this yields about
$\tau_{int}\simeq 10 fm/c$ \cite{lis05}, which is too short for $R_{1}$ and 
$R_{2}$ to reach equilibrium. Note, however, that this does
not apply to $R_{eq}$ which is born near local equilibrium at short 
times (in the present calculation, at t = 0 due to the
$\delta-$function intitial conditions),
and then spreads in time through diffusive interactions with other
particles at nearly the same rapidity. 
Although its variance does not fully attain the thermal
limit in the collisions investigated here, we refer
to $R_{eq}$ as the local equilibrium distribution since it comes 
very close to it.

Nonlinear effects are not considered here.
These account to some extent for the collective expansion of the
system in $y-$space, which is not included a priori in a statistical
treatment. In the linear model, the expansion is
treated through effective
diffusion coefficients $D_{y}^{eff}$ that are larger than the
theoretical values calculated from the dissipation-fluctuation theorem that 
normally relates $D_{y}$ and $\tau_{y}$ to each other \cite{wols99}.
One can then deduce the collective expansion velocities
from a comparison between data and theoretical result.  

The FPE can be solved analytically in the linear case  
with constant $D_{y}^{k}$. For net-baryon rapidity distributions,
the initial conditions are $\delta$-functions at the
beam rapidities $y_{b}=\pm y_{max}$. However, it has been shown that in
addition there exists a central (k=3, equilibrium) source at RHIC energies
which accounts for about 14{\%} of the net-proton yield in Au + Au
collisions at 200 AGeV \cite{wol03}, and is most likely related to
deconfinement. For d + Au, net-proton rapidity distributions are not
yet available.

For produced particles, the initial conditions are not uniquely
defined. Our previous experience with the Au + Au system
regarding both net baryons \cite{wol03}, and produced hadrons
\cite{biy04} favors a three-sources 
approach, with $\delta$-function initial conditions at the beam
rapidities, supplemented by a source centered at the equilibrium value
y$_{eq}$. This value is equal to zero
for symmetric systems, but for the asymmetric d + Au case its
deviation from zero according to (\ref{yeq}) is decisive 
in the description of particle production.

Physically, the particles in this source are expected to be generated
mostly from gluon-gluon collisions since only few valence quarks are
present in the midrapidity region at $\sqrt{s_{NN}}$ = 200 GeV
\cite{wol03}. Particle creation from a gluon-dominated source,
in addition to the sources related to the valence part of the 
nucleons, has also been proposed by Bialas and Czyz \cite{bia05}.
The final width of this source
corresponds to the local equilibrium temperature of the system which
may approximately be obtained from analyses of particle abundance ratios, plus
the broadening due to the collective expansion of the system.
Formally, the local equilibrium distribution is a solution
of (\ref{fpe}) with diffusion coefficient
$D_{y}^{3}$ = $D_{y}^{eq}$, and $\delta$-function initial condition at the
equilibrium value.
 
The PHOBOS-collaboration has analyzed their minimum-bias data 
successfully using a triple Gaussian fit
\cite{bbb04}. This is consistent with our analytical
three-sources approach, although additional
contributions to particle production have been proposed. 
Beyond the precise representation of the data, however,
the Relativistic Diffusion Model offers an analytical description of
the statistical equilibration during the collision and in particular, 
of the extent of the moving midrapidity source which is indicative
of a locally equilibrated parton plasma prior to hadronization.

With $\delta-$function initial conditions for the Au-like source (1),
the d-like source (2) and the equilibrium source (eq), we obtain 
exact analytical diffusion-model solutions as an incoherent
superposition of the distribution functions $R_{k}(y,t)$ because the
differential equation is linear. The three individual distributions
are Gaussians with mean values
\begin{equation}
<y_{1,2}(t)>=y_{eq}[1-exp(-t/\tau_{y})] \mp y_{max}\exp{(-t/\tau_{y})}
\label{mean}
\end{equation}
for the sources (1) and (2), and $y_{eq}$ for the
moving equilibrium
source. Hence, all three mean values attain y$_{eq}(b)$ as determined
from (\ref{yeq}) for t$\rightarrow \infty$, whereas for short times
the mean rapidities are smaller than, but close to the Au- and
d-like values in the sources 1 and 2. The variances are
\begin{equation}
\sigma_{1,2,eq}^{2}(t)=D_{y}^{1,2,eq}\tau_{y}[1-\exp(-2t/\tau_{y})].
\label{var}
\end{equation}

The charged-particle distribution in rapidity space is then obtained
as incoherent 
superposition of nonequilibrium and local equilibrium solutions of
 (\ref{fpe}) 
\begin{equation}
\frac{dN_{ch}(y,t=\tau_{int})}{dy}=N_{ch}^{1}R_{1}(y,\tau_{int})+
N_{ch}^{2}R_{2}(y,\tau_{int})
+N_{ch}^{eq}R_{eq}^{loc}(y,\tau_{int})
\label{normloc1}
\end{equation}
with the interaction time $\tau_{int}$ (total integration time of the
differential equation), and the partial distributions (k=1,2,eq)
\begin{equation}
R_{k}(y,\tau_{int})=\frac{1}{\sqrt(2\pi \sigma^{2}_{k}(\tau_{int}))}á
exp\Bigl[-\frac{(y-<y_{k}(\tau_{int})>)^{2}}{2\sigma_{k}^{2}(\tau_{int})}\Bigr].
\label{gauss}
\end{equation}

The incoherent sum of these distributions differs decisively from a 
single Gaussian that is sometimes taken to model pseudorapiditry
distributions for produced particles (together with the Jacobian
transformation that would generate the dip seen in the data at
midrapidity for symmetric systems, cf. sect.3).

In the present work, the integration is 
stopped at the value of $\tau_{int}/\tau_{y}$ that produces the
minimum $\chi^{2}$ with respect to the data and hence, the
explicit value of $\tau_{int}$ is not needed as an input. 
The result for central collisions is $\tau_{int}/\tau_{y} 
\simeq 0.4$ for d + Au, and $\tau_{int}/\tau_{y} 
\simeq 0.46$ for Au + Au. As the time evolution parameter in the 
actual numerical calculation we take $p=(1-exp(-2t/\tau_{y}))$,
and the corresponding values are p= 0.55 for d + Au, and 0.6
for Au + Au.

The average numbers of charged particles in
the target- and projectile-like regions $N_{ch}^{1,2}$ are 
proportional to the respective
numbers of participants $N_{1,2}$,
\begin{equation}
N_{ch}^{1,2}=N_{1,2}\frac{(N_{ch}^{tot}-N_{ch}^{eq})}{(N_{1}+N_{2})}
\label{nch}
\end{equation}
with the constraint $N_{ch}^{tot}$ = $N_{ch}^1$ + $N_{ch}^{2}$ +
$N_{ch}^{eq}$.
Here the total number of charged particles in each centrality bin
$N_{ch}^{tot}$ is determined from the data. The average number
of charged particles in the equilibrium source $N_{ch}^{eq}$ is a
free parameter that is optimized together with the variances
and $\tau_{int}/\tau_{y}$ in a $\chi^{2}$-fit of the data
using the CERN minuit-code \cite{jam81}. With known $\tau_{int}$, 
including its dependence on centrality, one could then 
determine $\tau_y$ and $D_y$, but this is beyond the scope of the present work.

\section{Application to d + Au collisions}

The time evolution of the resulting RDM-solutions is shown in 
Fig.1 for central collisions of d + Au at $\sqrt{s_{NN}}$ = 200 GeV
for short values of $\tau_{int}/\tau_{y}=0.005$ (p=0.01),
and large values $\tau_{int}/\tau_{y}=2.3$ (p=0.99).
In the latter case, the system
is already very close to statistical equilibrium in
pseudorapidity space, as is evident from the distribution functions 
shown in the lower part of Fig.1, which are almost symmetric with respect to the 
equilibrium value. In the actual collision, the system
remains between these two extreme cases, and in particular, it
remains far from the equilibrium situation, because strong
interaction stops long before this situation is approached, see Fig.2.

We present the results in pseudorapidity 
space $\eta=-ln[tan(\theta / 2)]$ since particle 
identification was not available. The conversion from $y-$ to $\eta-$
space of the rapidity density
\begin{equation}
\frac{dN}{d\eta}=\frac{dN}{dy}\frac{dy}{d\eta}=\frac{p}{E}\frac{dN}{dy}=
J(\eta,\langle m\rangle/\langle p_{T}\rangle)\frac{dN}{dy} 
\label{deta}
\end{equation}
is performed through the Jacobian
\begin{equation}
J(\eta,\langle m\rangle/\langle p_{T}\rangle) 
 = \cosh({\eta})\cdot [1+(\langle m\rangle/\langle p_{T}\rangle)^{2}
+\sinh^{2}(\eta)]^{-1/2}.
\label{jac}
\end{equation}
Here we approximate the average mass $<m>$ of produced charged hadrons in the
central region by the pion mass $m_{\pi}$, and use a
mean transverse momentum $<p_{T}>$ = 0.4 GeV/c. In the
Au-like region, the average mass is larger due to the
participant protons, but since their number $Z_{1}< 5.41$ is small compared to the 
number of produced charged hadrons in the d + Au system, the
increase above the pion mass remains small: 
$<m>\approx m_{p}\cdot Z_{1}/N_{ch}^{1} + m_{\pi}\cdot
(N_{ch}^1-Z_{1})/N_{ch}^1 \approx 0.17  GeV$. 
This increase turns out to have a negligeable effect on the results
of the numerical optimization, where we use $<m>/<p_{T}>=0.45$ for
the Jacobian transformations in the three regions. For reasonable
deviations of the mean transverse momentum from 0.4 GeV/c, the
results remain consistent with the data within the experimental error 
bars.

The result of the RDM calculation is shown in Fig. 2
for five collision centralities of d + Au, and minimum-bias,
and compared to recent PHOBOS data \cite{bbb05,bbb04}.
In case of central collisions, the charged-particle yield is
dominated by hadrons produced from the Au-like source, but there
is a sizeable equilibrium source that is more important
than the d-like contribution. This thermalized source is moving since
y$_{eq}$ has a negative value for d + Au, whereas it is zero
for symmetric systems.

The equilibrium source in the light and asymmetric d + Au system
is found to contain only 19\% of the produced charged hadrons in 
central collisions.
The total particle number and the particles created from
the Au-like source decrease almost linearly with increasing impact
parameter, but the magnitude of the equilibrium source is found to be 
roughly independent of centrality \cite{wbs05}. As a consequence, particle production
in the equilibrium source is relatively more important
in peripheral collisions. The variance of the central source
lies for sufficiently small impact parameters between the values for
the Au- and d-like sources \cite{wbs05}. In the equilibrium source,
a statistical descrition of particle production in terms of
a temperature and a chemical potential is meaningful. This is, 
however, not the case for the nonequilibrium fractions of the 
distribution function.

The minimization procedure yields precise results
so that reliable values for the relative importance of the
three sources for particle production can be determined, Table 1.
Here the average impact parameters $<b_{j}>$ for the five centrality 
cuts $j$ are determined
according to
\begin{equation}
<b_{j}>=\int b\sigma_{j}(b)db /\int\sigma_{j}(b)db
\label{imp}
\end{equation}
with the geometrical cross sections $\sigma_{j}(b)$. 
In a sharp-cutoff model with limiting impact parameters $b_{1},
b_{2}$ in each centrality bin ${j}$, this is
\begin{equation}
<b_{j}>=\frac{2}{3}\Bigl(\frac{b_{2}^{3}-b_{1}^{3}}{b_{2}^{2}-b_{1}^{2}}\Bigr)_{j}. 
\label{bimp}
\end{equation}
Whereas
the total particle number and the particles created from
the Au-like source decrease almost linearly with increasing impact
parameter, the magnitude of the equilibrium source is roughly
independent of centrality. As a consequence, particle production
in the equilibrium source is relatively more important
in peripheral collisions. The variance of the central source
lies for sufficiently small impact parameters between the values for
the Au- and d-like sources.

The rapidity relaxation times and diffusion coefficients
can also be obtained from (\ref{mean}),(\ref{var}),
but this requires an independent information about the interaction
times. A small discrepancy in case of the most
peripheral collisions (80-100\%) is a consequence of 
the three straggling data points in the region -4 $<\eta<-3$.

The observed shift of the distributions towards
the Au-like region in more central collisions, and the steeper slope 
in the deuteron direction as compared to the gold direction
appear in the Relativistic Diffusion Model as a
consequence of the gradual (incomplete) approach to equilibrium.
The dependence of the shape and the absolute magnitudes on centrality 
are particularly evident in Fig.3 where the pseudorapidity 
distributions for all centralities are shown in a single plot 
in comparison with the PHOBOS data.

Given the structure of the underlying differential equation
that we use to model the equilibration,
together with the initial conditions
and the constraints imposed by Eqs. ($\ref{yeq})$ and ($\ref{nch}$),
there is no room for substantial modifications of this result.
In particular, changes in the impact-parameter dependence of the mean 
values in (\ref{mean}) that are not in accordance with 
(\ref{yeq}) vitiate the precise agreement with the data.

\section{Application to Au + Au}

The three-sources RDM has previously been applied to the 
Au + Au system at various incident energies for net protons
\cite{wol03}, and for produced charged hadrons \cite{biy04}.
For net protons, the number of particles contained
in the midrapidity source can be determined rather
accurately, whereas this is not the case for produced
charged hadrons. This is due to the uncertainty in
the initial conditions ($\delta-$functions
are clearly correct for the participant protons, but they
remain ambiguous for produced particles), and also due to
the symmetry of the system, which does not permit unique results
of a $\chi^{2}-$fit of the data.

A previous result \cite{biy04} for Au + Au in the 
three-sources-RDM shows indeed that the size of the equilibrium source for 
particle production at a given centrality can not be 
determined uniquely, and may be 
different in the heavy
system \cite{biy04} at the same energy as compared to d + Au. Here we
fix the particle content in the midrapidity source at a given value
of $<b>/b_{max}$ 
to approximately the same value as in the d + Au case,
where the result is given from fitting the analytical solutions to 
the data (see Tab.1). Here $<b>$ is determined for a given centrality
as described in the previous section.

The result for central Au + Au-collisions 
at $\sqrt{s_{NN}}$ = 200 GeV is shown in Fig.4 (upper frame)
together with PHOBOS data \cite{nou02}. The time evolution of the
RDM-solutions can be seen by comparing short-time solutions (middle 
frame for short values of $\tau_{int}/\tau_{y}=0.005$, or p=0.01)
with the solutions for large times (lower frame for
 $\tau_{int}/\tau_{y}=2.3$, or p=0.99).
In the latter case, the system
is again very close to statistical equilibrium in
pseudorapidity space. The actual collision with
$\tau_{int}/\tau_{y}=0.46$, or p=0.6, (top frame)
remains between these two extreme cases.

The comparison with PHOBOS data for various centralities
can be seen in Fig.5, where the individual partial distributions
are also shown. For each centrality, the percentage of
produced charged hadrons is taken approximately from the d + Au results.
It rises for more peripheral collisions, because the number
of charged hadrons produced from nucleon-nucleon collisions in the
target- and projectile-like region of pseudorapidity space falls
more strongly than the overall number of produced hadrons.
However, the formation of an equilibrated quark-gluon plasma
in the local equilibrium region prior to hadronization can probably
only be expected for central collisions, since it requires
high excitation and density.

\section{Conclusion}

To conclude, we have investigated 
charged-particle production in d + Au and Au + Au collisions at
$\sqrt{s_{NN}}$= 200 GeV as function of centrality
within the framework of an analytically soluble three-sources 
model. Excellent agreement with recent PHOBOS pseudorapidity
distributions has been obtained, and from a $\chi^{2}$-minimization we have 
determined the diffusion-model parameters
very accurately. 

For central d + Au collisions, a fraction of only 
19\% of the produced particles arises from the locally equilibrated
midrapidity source. Although this fraction increases towards more
peripheral collisions, the formation of a thermalized parton
plasma prior to hadronization can probably only be expected for
more central collisions.

The d + Au results show clearly that only the midrapidity part of the
distribution function comes very close to thermal equilibrium, whereas the
interaction time is too short for the d- and Au-like parts 
to attain the thermal limit. The same is true for the heavy Au + Au 
system at the same energy, but there the precise fraction of
particles produced in the equilibrium source is more
difficult to determine due to the symmetry of the problem.

The relativistic systems can thus be seen to be on their way
towards statistical equilibrium. However, due to the dynamical 
evolution both the asymmetric and the symmetric system remain
far from reaching thermodynamic equilibrium, which is closely 
approached only by the hadrons created from the central source that
is mostly due to gluon-gluon collisions.\\

{\bf Acknowledgement} One of the authors (GW) acknowledges 
the hospitality of the Faculty
of Sciences at Shinshu University, and financial support by
the Japan Society for the Promotion of Science (JSPS)
which are due to contacts at RCNP (Osaka University) and
the Yukawa Institute of Theoretical Physics (YITP) at Kyoto University.

\newpage
\rm
Table 1. Produced charged hadrons as functions of
centrality in d + Au  
collisions at $\sqrt{s_{NN}}$ =
200 GeV, y$_{b}=\pm$ 5.36 in the Relativistic Diffusion Model.
The average impact parameter for each centrality bin is $<b>$,
the corresponding equilibrium value of the rapidity
is $y_{eq}$, the variance of the central source in
$y-$space is $\sigma_{eq}^{2}$; $<N_{1,2}>$\cite{bbb05}
are the respective average 
numbers of participants.
The number of produced charged particles 
is $N_{ch}^{1,2}$ for the sources 1 and 2 and $N_{ch}^{eq}$ for the equilibrium
source, the percentage of
charged particles produced in the thermalized source is $n_{ch}^{eq}$.
\\[1.5cm]
\begin{tabular}{lccccccccccc}
\hline
\hline
$Centrality  (\%)$ &$<b>(fm)$& $y_{eq}$&
$\sigma_{eq}^{2}$&$<N_{1}>$&$<N_{2}>$
&$N_{ch}^{1}$&$N_{ch}^{2}$&$N_{ch}^{eq}$&$n_{ch}^{eq}$(\%)&\\
\hline
0-20 &2.53&- 0.944& 3.99 &13.5&2&   131 &  19 &35&19\cr
20-40 &4.63& - 0.760&3.95 &8.9&1.9&  78 &  17 &31&25\cr
40-60 &5.99&- 0.564& 5.70 &5.4&1.7&  33 &  11 &38&46\cr
60-80 &7.10& - 0.347&7.44 &2.9&1.4&  9 &  5 &35&71\cr
80-100 &8.05&- 0.169& 6.89 &1.6&1.1&   2 &  2 &24&86\cr
\hline
min. bias&5.66&- 0.664&4.04 &6.6&1.7&  56 & 15 & 21 & 23 \cr 
\hline
\hline
\end{tabular}
\newpage
\Large\bf
Figure captions
\normalsize\rm
\begin{description}
\item[Fig. 1]
Time evolution of the analytical solutions in the three-sources
Relativistic Diffusion Model (RDM) for minimum-bias d + Au 
collisions at $\sqrt{s_{NN}}$ = 200 GeV. The curves in the upper frame show
the three partial pseudorapidity distributions $(dN_{ch}/d\eta)_{k}$
and their incoherent sum at short times
$\tau_{int}/\tau_{y}=0.005$ corresponding to p=0.01. In the lower 
frame, $\tau_{int}/\tau_{y}=2.3$ corresponding to p=0.99 displays
the solutions for large times where they come close to statistical 
equilibrium. Strong interaction stops long before this situation is 
reached, see Fig.2.
\item[Fig. 2]
Calculated pseudorapidity distributions of charged hadrons in
d + Au collisions at $\sqrt{s_{NN}}$ = 200 GeV for five different
collision centralities, and minimum-bias in comparison with
PHOBOS data \cite{bbb05,bbb04}. The analytical RDM-solutions are
optimized in a fit to the data. The corresponding 
minimum $\chi^{2}$-values (top left to bottom right) are 4.7, 5.9,
2.4, 1.7, 1.9, 2.1. Au-like, d-like, and central partial distributions
are shown for each centrality. Only the midrapidity part comes close 
to local equilibrium.
\item[Fig. 3]
Calculated pseudorapidity distributions of charged particles from
d + Au collisions at $\sqrt{s_{NN}}$ = 200 GeV for five different
collision centralities, and minimum-bias in comparison with
PHOBOS data \cite{bbb05,bbb04}. The steeper slope in the deuteron 
direction is due to the nonequilibrium properties of the system.
\item[Fig. 4]
Time evolution of the analytical solutions in the three-sources
Relativistic Diffusion Model (RDM) as in Fig.1, but for central 
(0-6\%) Au + Au 
collisions at $\sqrt{s_{NN}}$ = 200 GeV. The curves in the middle frame show
the three partial pseudorapidity distributions at short times
$\tau_{int}/\tau_{y}=0.005$ corresponding to p=0.01. In the lower 
frame, the distributions are close to equilibrium
with $\tau_{int}/\tau_{y}=2.3$ (p=0.99). The upper
frame gives the comparison with PHOBOS data \cite{nou02}; here,
$\tau_{int}/\tau_{y}=0.46$ (p=0.6).
\item[Fig. 5]
Calculated pseudorapidity distributions of charged hadrons in
Au + Au collisions at $\sqrt{s_{NN}}$ = 200 GeV for six different
collision centralities in comparison with
PHOBOS data \cite{bb105,nou02}. The relative strength of 
the midrapidity source at each centrality
is chosen here in analogy to d + Au, cf. text.
The analytical RDM-solutions are
optimized in a fit to the data. The corresponding 
minimum $\chi^{2}$-values (top left to bottom right) are 1.2, 1.0,
0.95, 0.67, 0.53, 0.70. Only the midrapidity part comes close to 
local equilibrium. Its relative importance increases 
with decreasing centrality.
\end{description}
\newpage
\vspace{1cm}
\includegraphics[width=12cm]{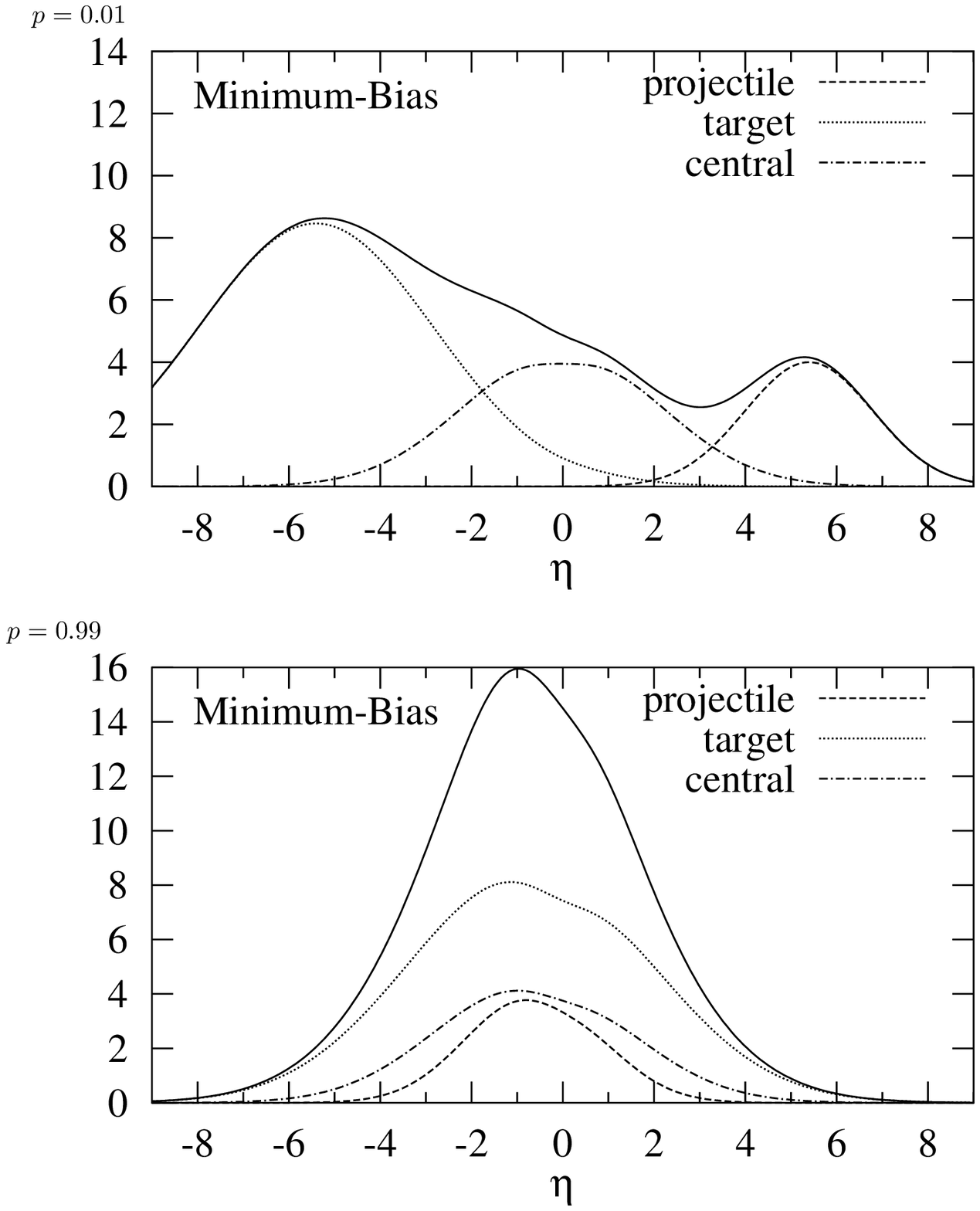}
\newpage
\vspace{1cm}
\includegraphics[width=15.2cm]{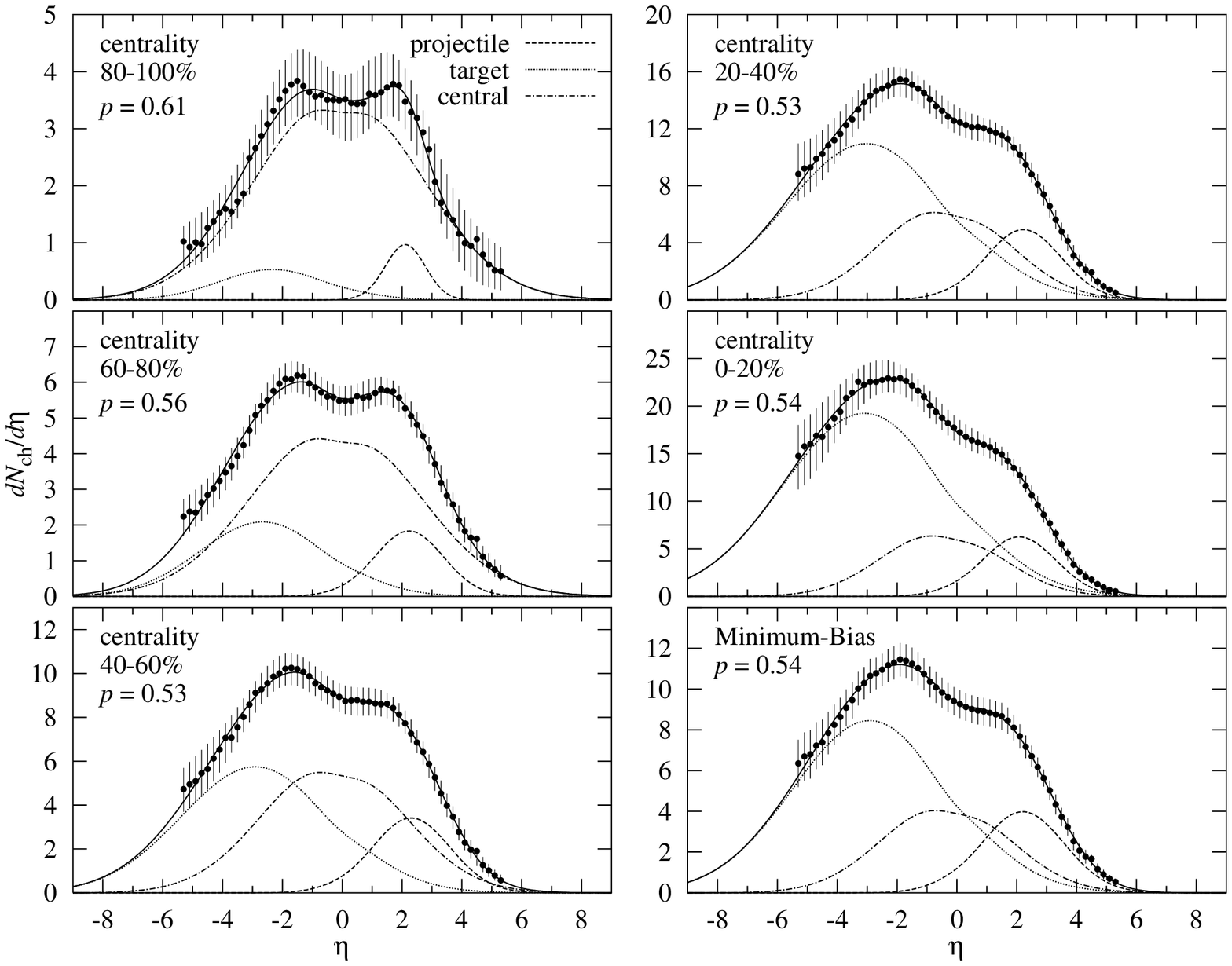}
\newpage
\vspace{1cm}
\includegraphics[width=16cm]{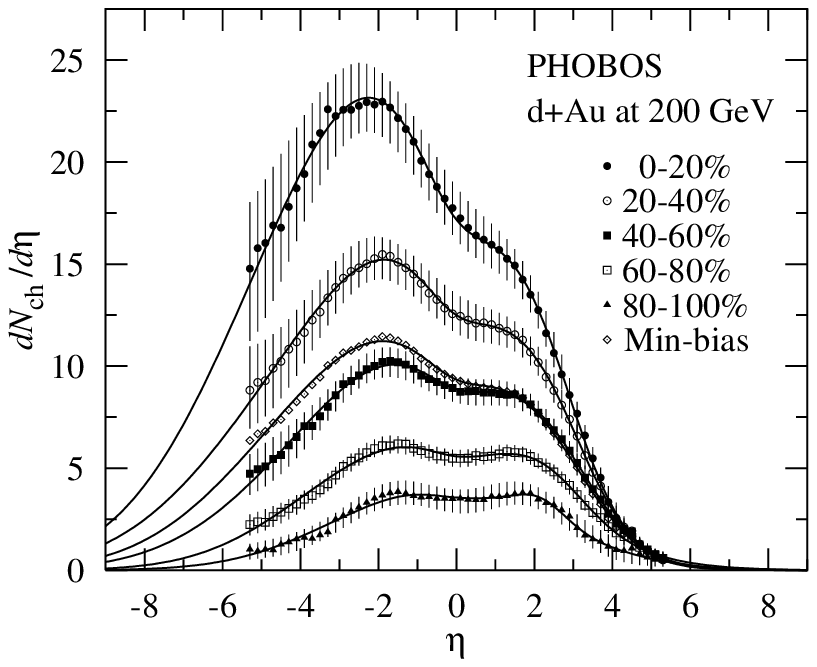}
\includegraphics[width=12cm]{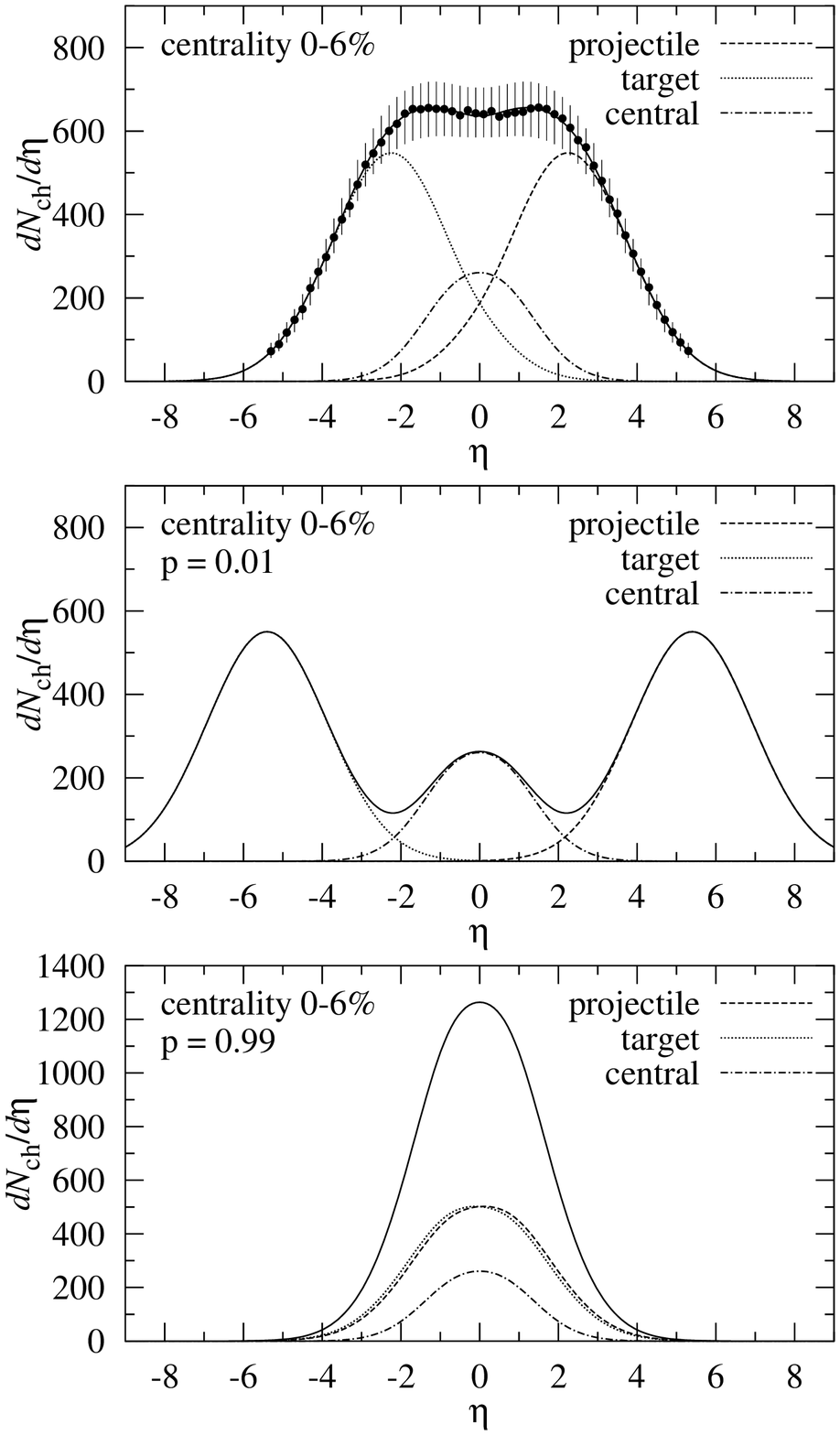}
\includegraphics[width=15.2cm]{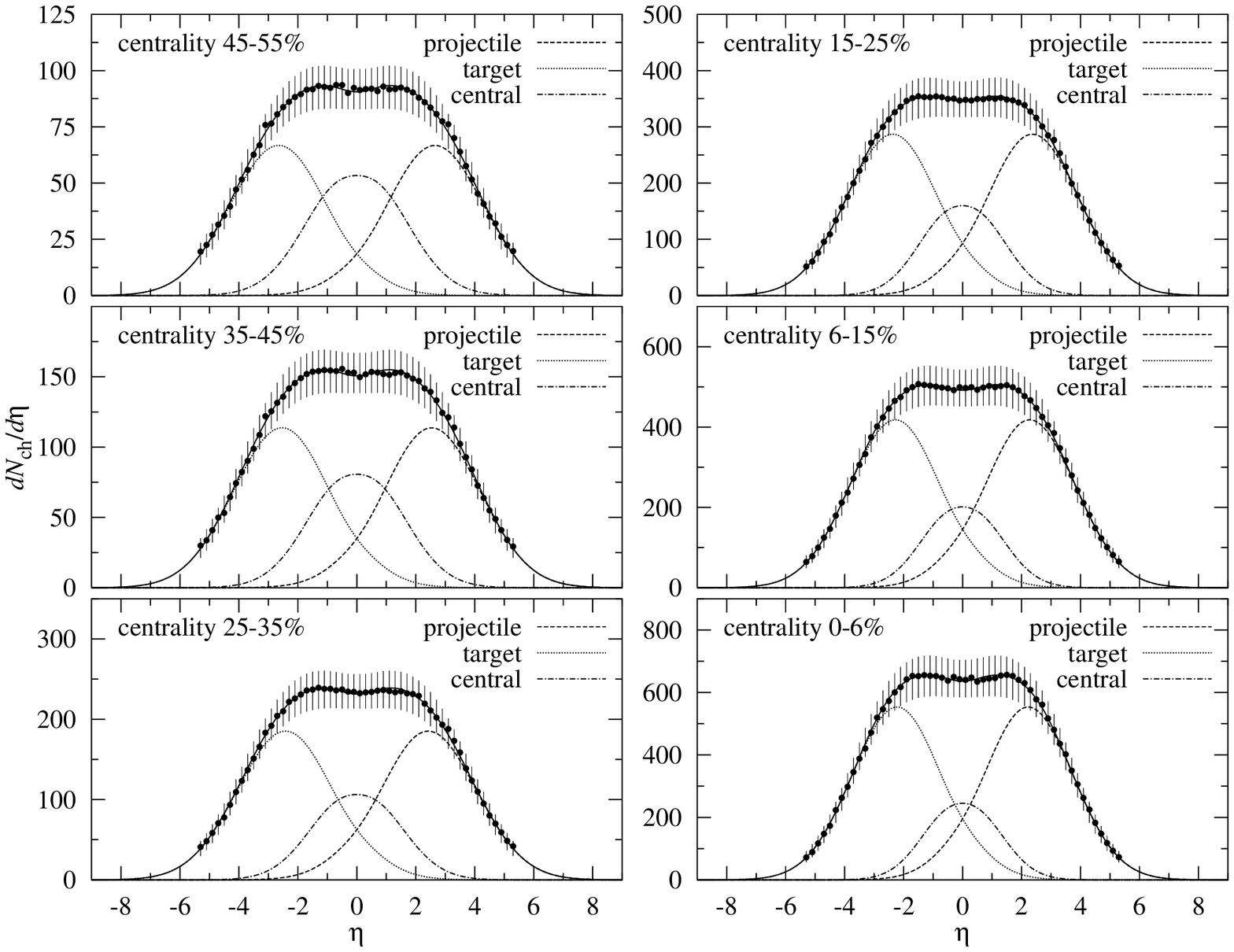}

\newpage
\begin{thebibliography}{10}
\bibitem{bbb05}B.B. Back, et al.,
Phys. Rev. C {\bf 72}, 031901 (2005).
\bibitem{bb105}B.B. Back, et al., nucl-ex/0509034,
submitted to Phys. Rev. C.
\bibitem{nou02}R. Nouicer et al., nucl-ex/0208003,
Proc. XXXVII Rencontres de Moriond (2002).
\bibitem{wbs05}G. Wolschin, M. Biyajima, T. Mizoguchi, and N. Suzuki,
\\ Phys. Lett. B {\bf 633}, 38 (2006). 
\bibitem{fok14}A.D. Fokker, Ann. Phys. (Leipzig) {\bf 43}, 810 
(1914);\\
M. Planck, Sitzber. Preu{\ss}. Akad. Wiss. p. 324 (1917).
\bibitem{kam97}N.G. van Kampen, Phys. Bl. {\bf 53}, 1012 (1997).
\bibitem{ray91}Lord Rayleigh, Phil. Mag. {\bf 32}, 424 
(1891).
\bibitem{ein05}A. Einstein, Ann. Phys. (Leipzig) {\bf 17}, 549 
(1905); {\bf 19}, 289 (1906); {\bf 19}, 371 (1906).
\bibitem{uhl30}G.E. Uhlenbeck and L.S. Ornstein,
Phys. Rev. {\bf 36}, 823 (1930).
\bibitem{rhic05}I. Arsene et al., Nucl. Phys. A {\bf 757}, 1 (2005).
\bibitem{wol99}G. Wolschin, Eur. Phys. J. A {\bf 5}, 85 (1999).
\bibitem{biy02}M. Biyajima, M. Ide, T. Mizoguchi, and N. Suzuki,\\
Prog. Theor. Phys. {\bf 108}, 559 (2002); {\bf 109}, 151 (2003).
\bibitem{wol03}G. Wolschin, Phys. Lett. B {\bf 569}, 67 (2003);\\
Phys. Rev. C {\bf 69}, 024906 (2004); hep-ph/0502123.
\bibitem{bha53}H.J. Bhabha, Proc. Roy. Soc. (London) A {\bf 219}, 293 (1953).
\bibitem{nag84}S. Nagamiya and M. Gyulassy, Adv. Nucl. Phys. {\bf 13}, 201
(1984).
\bibitem{lis05}M. Lisa, S. Pratt, R. Soltz, and U. Wiedemann,
\\Ann. Rev. Nucl. Part. Sci. {\bf 55}, 357 (2005).
\bibitem{ryb03}M. Rybczy\'nski, Z. W{\l}odarczyk, and G. Wilk,\\ Nucl. 
Phys. B (Proc. Suppl.) {\bf 122}, 325 (2003).
\bibitem{wols99}G. Wolschin, Europhys. Lett. {\bf 47}, 30 (1999).
\bibitem{biy04}M. Biyajima, M. Ide, M. Kaneyama, T. Mizoguchi, and 
N. Suzuki,\\
Prog. Theor. Phys. Suppl. {\bf 153}, 344 (2004).
\bibitem{bia05}A. Bialas and W. Czyz, Acta Phys. Polon. B {\bf 36}, 
905 (2005). 
\bibitem{bbb04}B.B. Back, et al., Phys. Rev. Lett. {\bf 93}, 082301 (2004).
\bibitem{jam81}F. James, CERN Report 81-03 (1981).
\end{thebibliography}
\end{document}